\documentstyle[psfig,editedvolume]{crckapb} 
\def\lesssim{\mathrel{\hbox{\rlap{\hbox{\lower4pt\hbox{$\sim$}}}\hbox{$<$}}}}
\def\gtrsim{\mathrel{\hbox{\rlap{\hbox{\lower4pt\hbox{$\sim$}}}\hbox{$>$}}}}
\let\simlt\lesssim\let\simgt\gtrsim
\let\ctyr\citeyear
\def\kms{\hbox{${\rm{}km\;s}^{-1}$}}
\def\apj{ApJ}\def\aap{A\&A}\def\mnras{MNRAS}
\begin{opening}
\title{UNIQUE WHITE DWARFS ACCOMPANYING RECYCLED PULSARS}
\author{M. H. van Kerkwijk}
\institute{Palomar Observatory\\
           California Institute of Technology, mail stop 105-24\\
           Pasadena, CA 91125, USA}
\end{opening}
\begin{document}

\section{Introduction and Summary}

Knowledge of the properties of the white-dwarf companions of radio
pulsars can provide unique constraints on the characteristics and
evolution of these binaries, as well as on those of its constituents.
Many white-dwarf properties can be determined accurately from their
spectra, but for the very faint pulsar companions spectroscopy has
only become feasible with the advent of large telescopes like the
Keck.  Here, I introduce the two classes of pulsar, white-dwarf
binaries, and describe for each what we have learned from a specific
system, PSR~J1012+5307 and PSR~B0655+64, respectively, summarising
what has been done (Van Kerkwijk \& Kulkarni \ctyr{vkerk:95}; Van
Kerkwijk, Bergeron, \& Kulkarni \ctyr{vkerbk:96}; hereafter Papers I
\& II), presenting new results, and discussing what the future may
hold.

Briefly, for the companion of PSR~J1012+5307 we find a DA spectrum,
and infer a mass of $\sim\!0.16\;M_\odot$, the lowest among all
spectroscopically identified white dwarfs.  Combined with a
radial-velocity orbit, a neutron-star mass between 1.5 and
$3.2\;M_\odot$ (95\% confidence) is derived.  The companion of
PSR~B0655+64 shows strong Swan C$_2$ bands, i.e., it is a DQ star.
Unlike anything reported for other DQs, however, it shows variations
in strength of the bands by a factor two.  Most likely, the variations
are periodic, with a period of $\sim\!9.7\;$h.  This is substantially
shorter than the $1\;$day orbital period, which can likely be
understood in terms of its past evolution.

\section{PSR~J1012+5307\label{sec:psrj1012}}

PSR~J1012+5307 is a member of the largest group of binary pulsars,
those with low-mass helium white dwarf companions.  These systems have
presumably descended from the low-mass X-ray binaries, in which mass
is transferred onto a neutron star from a less massive companion.  The
mass transfer in these systems is stable and relatively well
understood, and a number of predictions can be made (for a review of
binary evolution involving neutron stars, see Bhattacharya \& Van den
Heuvel \ctyr{bhatvdh:91}).  First, the neutron star will likely have
accreted a substantial fraction, if not all, of the up to
$0.7\;M_\odot$ lost by the companion (e.g., Van den Heuvel \&
Bitzaraki \ctyr{vdheb:95}).  Hence, one expects the neutron star to
have increased substantially in mass, and to be spun up\footnote{These
pulsars are called ``recycled'' because the spin-up process allows the
radio-pulsar mechanism to work again after the mass transfer ceases.}.
For reasons not quite understood, the magnetic field seems to decay at
the same time, perhaps by being quite literally buried (Romani
\ctyr{roma:90}).  If the neutron stars indeed have masses increased to
$\simgt\!2\;M_\odot$ (assuming they started with the ``canonical''
$1.4\;M_\odot$; for a recent census, see Van Kerkwijk et al.\
\ctyr{vkervpz:95}), this would be very interesting, as it would
strongly constrain the equation of state (EOS) at supra-nuclear
densities (e.g., Cook, Shapiro, \& Teukolsky \ctyr{cookst:94}): for
softer EOS, like the one recently proposed by Brown \& Bethe
(\ctyr{browb:94}), such a massive neutron star would collapse into a
black hole.  The only system for which a neutron-star mass estimate is
available, is PSR~B1855+09, for which Kaspi, Taylor, \& Ryba
(\ctyr{kasptr:94}) found $M_{\rm{}NS}=1.50^{+0.26}_{-0.14}\;M_\odot$
(68\% confidence).  As yet, the uncertainty is too high to allow one
to draw a strong conclusion.

For the white dwarf, one predicts that it will have a helium core,
since the companion never reached helium ignition.  Furthermore, for
the systems with orbital periods $\simgt\!10\;$d, there should be a
relation between the orbital period and the white-dwarf mass (as a
consequence of the core-mass, radius relation for giants; Refsdal \&
Weigert \ctyr{refsw:71}; most recently for pulsar binaries, Rappaport
et al.\ \ctyr{rapp&a:95}), as well as a statistical relation between
the orbital period and the eccentricity (Phinney \ctyr{phin:92}).  The
latter depends only on the assumption of having a Roche-lobe filling
giant with a convective envelope at the end of the evolution, and has
been confirmed (ibid.).  The orbital-period, mass relation has only
been verified at the short-period end, using (again) PSR~B1855+09, for
which Kaspi et al.\ (\ctyr{kasptr:94}) determined an accurate
companion mass from Shapiro delay in the pulse arrival times.  At the
long period end, we hope to obtain an additional constraint using
PSR~B0820+02, which has a DA companion (Paper~I).

PSR~J1012+5307 is a recently discovered $5.26\;$ms pulsar, which is in
a $0.60\;$d orbit with a very low-mass companion (Nicastro et al.\
\ctyr{nica&a:95}).  This system was deemed especially interesting as,
given its small orbital period, it seemed possible to determine the
radial-velocity amplitude and thus the mass ratio.  Combined with the
white-dwarf mass, this would give the mass of the neutron star.

Fortunately, the optical counterpart has $V=19.6$ (Lorimer et al.\
(\ctyr{lori&a:95}), making it the brightest pulsar companion currently
known.  Partly, it is so bright because it is relatively hot, with a
colour-temperature of $T_{\rm{}BB}\simeq9400\;$K.  This indicates a
cooling age of only a couple $10^8\;$yr, much shorter than the pulsar
spin-down age of $7\,10^9\;$yr.  Since the pulsar presumably started
spinning down at the time the white dwarf was formed, this is
puzzling.  It probably indicates that the pulsar did not start
spinning at much shorter periods, as implicitly assumed in calculating
the spin-down age, and as would have been thought based on simplistic
models for the mass transfer.  It could also be, however, that these
low-mass white-dwarf have some residual hydrogen burning for quite a
while after losing the red-giant envelope, and that the cooling-age
estimate is wrong (Alberts et al.\ \ctyr{albe&a:96}).

\begin{figure}
\centerline{\hbox{\psfig{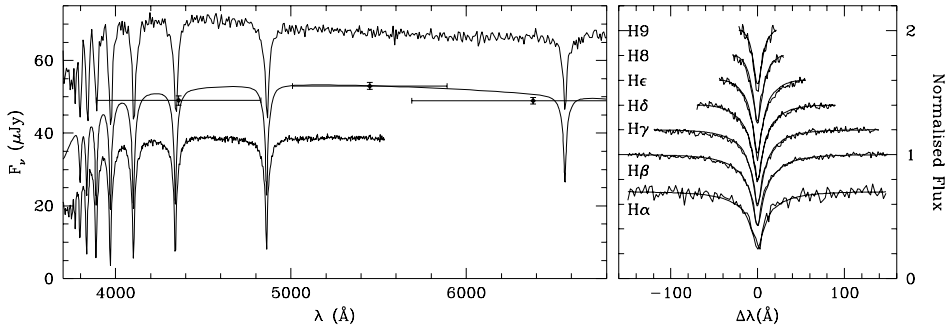}}}
\caption[]{The spectrum of the white-dwarf companion of PSR J1012+5307
(taken from Paper~II).  Shown in the left-hand panel are a
$10\;$\AA\ resolution classification spectrum (top curve; offset by
$15\;\mu$Jy) and the average of eight $4\;$\AA\ resolution spectra
(bottom curve; offset by $-15\;\mu$Jy).  Also shown are the broad-band
fluxes of Lorimer et al.\ (1995) and the best-fit pure Hydrogen model
spectrum.  The latter was derived from a fit to the profiles of
H$\beta$ up to H8, and has $T_{\rm{}eff}=8550\pm25\;$K and
$\log{}g=6.75\pm0.07$.  In the right-hand panel, the observed line
profiles, including those of H$\alpha$ and H9, are shown with the
modeled ones superposed.\label{fig:j1012model}}
\end{figure}

We found that the companion was a DA star, showing H$\alpha$ up to H12
(Paper~II; see Fig.~\ref{fig:j1012model}).  From a model-atmosphere
fit (Fig.~\ref{fig:j1012model}), we find $T_{\rm{}eff}=8550\pm25\;$K
and $\log{}g=6.75\pm0.07$ (cgs units).  To infer the mass, we need a
mass-radius relation.  Unfortunately, for these very low-mass helium
white dwarfs, this is not well known.  Using the Hamada-Salpeter
zero-temperature relation, with an approximate finite-temperature
correction based on models of Wood (\ctyr{wood:95}), we find
$M_{\rm{}WD}=0.16\pm0.02\;M_\odot$, the lowest among all
spectroscopically identified white dwarfs.

We also measured radial velocities, and found a radial-velocity
amplitude $K_{\rm{}WD}=280\pm15\;\kms$, leading to a mass ratio
$M_{\rm{}NS}/M_{\rm{}WD}=13.3\pm0.7$.  Combined with the white-dwarf
mass and the pulsar mass function, we infer that with 95\% confidence
$1.5<M_{\rm{}NS}/M_\odot<3.2$ (Paper~II).

This determination is not yet accurate enough to constrain the
equation of state, or to test evolutionary theory, but it does show
that further study may well prove fruitful.  It will be relatively
straightforward to improve the accuracy of the radial-velocity
amplitude and thus the mass ratio, which might already lead to an
interesting constraint on the mass of the neutron star.  It will be
less easy to improve the estimate of the white-dwarf mass, because of
the uncertainties in the mass-radius relation for these very low-mass
white dwarfs, as well as the possible presence of helium in the
atmosphere.  If helium is present, the true surface gravity---and thus
the inferred mass---will be lower (Bergeron, Wesemael, \& Fontaine
\ctyr{bergwf:91}; Reid \ctyr{reid:96} for an observational
indication).

The pulsar is relatively nearby and bright, however, and it may well
be possible to derive an accurate distance using radio VLBI or
timing.  This would allow one to obtain a direct estimate of the
radius.  If this is the same as the predicted one
($0.028\pm0.002\;R_\odot$), it would give confidence in the result.
If it is not, one can either assume there is a problem with the
mass-radius relation, but not with helium pollution, and infer a mass
from the radius in combination with the observed surface gravity, or
one can assume that there is helium pollution, but that the
mass-radius relation is fine, and use that to derive a mass from the
radius.  Another possibility is to search carefully for Shapiro delay
in the pulse arrival times, which would give a constraint on a
combination of the white-dwarf mass and the inclination.  

\section{PSR~B0655+64\label{sec:psrb0655}}

The companion of PSR~B0655+64 has a mass $>\!0.67\;M_\odot$ (for a
$1.4\;M_\odot$ pulsar), and thus it must have been a relatively
massive star.  Most likely, the evolution has been similar to that
leading to double neutron-star binaries like the Hulse-Taylor pulsar,
with the system going through a phase as a wide high-mass X-ray
binary, followed by spiral-in during a common-envelope phase, leading
to its current $1.03\;$d orbital period (e.g., Bhattacharya \& Van den
Heuvel \ctyr{bhatvdh:91}).  Presumably, the helium core left was not
massive enough to form a second neutron star.  Since common-envelope
evolution is rather poorly understood, there are few predictions for
these systems, except that one expects a carbon-oxygen white dwarf.
Most likely, it will have a helium atmosphere, since all the hydrogen
left after the spiral-in will probably have disappeared during a
second stage of mass transfer when the star became a helium giant
(this is expected for low-mass helium stars; Paczynski \ctyr{pacz:71};
Habets \ctyr{habe:86}).

\begin{figure}
\centerline{\hbox{\psfig{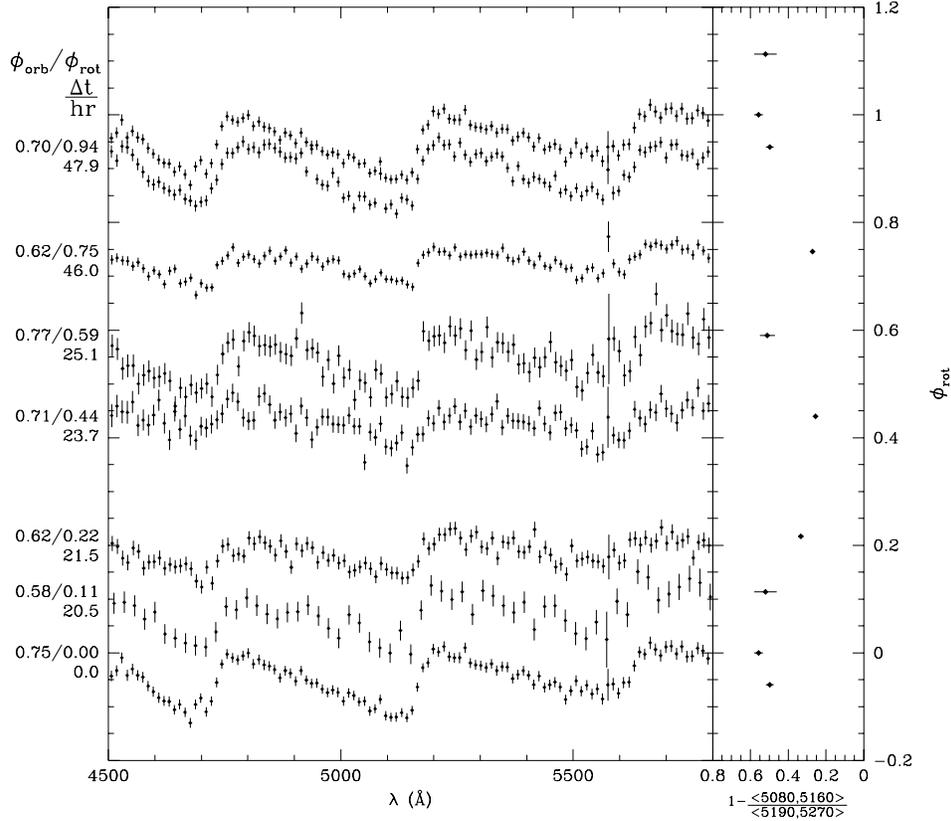}}}
\caption[]{Spectra of the companion of PSR~B0655+64.  In the
right-hand panel, a measure of the strength of the Swan bands is
given.  The numbers on the left indicate the time in hours since the
first observation, the orbital phase, and the rotation phase for a
period of $9.7\;$h.\label{fig:b0655}}
\end{figure}

The companion was identified by Kulkarni (\ctyr{kulk:86}).  It has
$V=22.2$ and $V-R=0.1$, indicating a temperature of about 6000 to
$9000\;$K.  First Keck spectra showed strong C$_2$ Swan bands
(Paper~I; see also Fig.~\ref{fig:b0655}), i.e., it is a DQ stars, with
a helium atmosphere sufficiently shallow and convective to allow trace
amounts of carbon to be dredged up (Pelletier et al.\
\ctyr{pell&a:86}).

Uniquely among DQ stars, a large variation in the strength of the Swan
bands was observed, by about a factor two in less than two hours.  In
Paper~I, this was interpreted as due to brighter and darker spots on
the white-dwarf surface, possibly related to the presence of a
magnetic fields (a locally higher magnetic field strength might lead
to a change in gas pressure and thus temperature, or in convective
efficiency).  Based on the speed and amplitude of the variation, it
was shown that if it was periodic, the period had to be $\simgt\!3$
and $\simlt\!12\;$h.  Thus, the modulation could not be orbital, but
was most likely due to the white dwarf rotation.  Spectra obtained in
November 1995 confirm this (Fig.~\ref{fig:b0655}), and a period of
$9.7\pm0.1\;$h is indicated.

In Paper~I, it was noted that if the star was rotating synchronously
with the orbit when it was a Roche-lobe filling helium giant, it would
have been spun up due to conservation of angular momentum when it
shrunk to form a white dwarf.  If the spin-up is mostly due to the
angular momentum contained in the remaining giant envelope, the
envelope mass must have been $\sim\!10^{-4.5}\;M_\odot$, interestingly
similar to the typical helium-layer masses inferred for DQ stars
(Pelletier et al.\ \ctyr{pell&a:86}; Weidemann \& Koester
\ctyr{weidk:95}; Dehner \& Kawaler \ctyr{dehnk:95}).

\acknowledgements I thank Yanqin Wu, Shri Kulkarni, and Pierre
Bergeron for useful discussions, and acknowledge support from a NASA
Hubble Fellowship.

\end{document}